\documentclass
[12pt,showpacs,preprintnumbers,nofootinbib,preprint,amsfonts,amssymb]{revtex4}%
\usepackage{amsmath}
\usepackage{graphicx}
\usepackage{bm}
\usepackage{amsfonts}
\usepackage{amssymb}
\usepackage{appendix}%
\begin{document}
\preprint{ }
\title{Dipole-Interacting Fermionic Dark Matters \\in Positron, Antiproton and Gamma-ray Channels }
\author{Jae Ho Heo}
\email{jaeheo1@gmail.com}
\author{C. S. Kim}
\email{cskim@yonsei.ac.kr}
\affiliation{Department of Physics and IPAP, Yonsei University, Seoul 120-479, Korea }

\begin{abstract}
\noindent Cosmic ray signals from dipole-interacting dark matter annihilation
are considered in the positron, antiproton and photon channels. The predicted
signals in the positron channel could nicely account for the excess of
positron fraction from Fermi LAT, PAMELA, HEAT and AMS-01 experiments for the
dark matter mass larger than 100 GeV with a boost (enhancement) factor of
$30-80$. No excess of antiproton over proton ratio at the experiments also
gives a severe restriction for this scenario. With the boost factors, the
predicted signals from Galactic halo and signals as mono-energetic gamma-ray
lines (monochromatic photons) for the region close to the Galactic center are
investigated. The gamma-ray excess of recent tentative analyses based on Fermi
LAT data and the potential probe of the monochromatic lines at a planned
experiment, AMS-02, are also considered.

\end{abstract}

\maketitle

\section{ Introduction}

The existence of dark matter (DM), as the invisible matter interacting by the
force of gravity, has been widely accepted by cosmological observations from
the experiments: Cosmic Microwave Background (CMB) \cite{eko09}, Galactic
Rotation Curves \cite{abo01}, Gravitational Lensing \cite{hho02} and Massive
Compact Halo Objects \cite{cal00}, and $etc$. About 83\% of the matter (around
23\% of the total energy density) in the universe is believed to be composed
of DM to account for the observations. However, the nature of DM is still
completely unknown despite decades of detection efforts. Many possible
explanations have been proposed. One of the alternative explanations from the
point of view of particle physics is that DM is composed of massive particles
and its interaction with ordinary matter is very weak.

Recently several DM models\footnote{Actually these models have been built to
explain the annual modulation signal from DAMA/NaI \cite{rbe03} and DAMA/LIBRA
\cite{rbe08} experiments with null results from other experiments (there is no
experimental evidence corroborating this signal yet). The scenario in Ref.
\cite{bfe10} is especially interesting, because the signal appears to be
electromagnetic energy deposit, not nuclear energy deposit, through the single
photon emission by the decay of the excited state.} (inelastic DM
\cite{bfe10,jing10}, asymmetric DM \cite{han10}, form factor DM \cite{bfel10})
with magnetic dipole interaction have been considered. DM in these models has
a few states and has no direct interaction with the photon. The candidate
particles could thus be stable and make up the invisible matter in our
universe. On the other hand, the direct interaction with photon through
magnetic dipole coupling has gotten some attention
\cite{heo09,mpo00,ksi04,vba11,wsc10,tba10,edn12} due to its plausibility. Most
of the works have concentrated on direct detections of DM. The main motivation
for the magnetic dipole interacting DM scenario is that the magnetic dipole
coupling can be sizable compared to other electromagnetic couplings, because
the magnetic dipole conserves the discrete symmetries like parity (P), time
reversal (T), and charge conjugation (C) or its combinations.

In this work, we consider cosmic ray signatures (indirect detections) of the
direct dipole-interacting DM with the shifted photon (hypercharge gauge boson
$B_{\mu}$). The cosmic ray signatures in the positron, antiproton and photon
channels are considered for the DM mass near the electroweak scale $(10-1000$
GeV$)$, essentially around 100 GeV. The dimension-5 operator which induces the
dipole interaction is $\overline{\psi}\sigma_{\mu\nu}\psi B^{\mu\nu}$, and it
may be expressed with photon and $Z$ boson in the standard model context since
the hypercharge gauge boson is a linear combination of photon and $Z$ boson
with Weinberg angle $\theta_{W}$. The relevant effective Lagrangian is given
by%
\begin{equation}
\mathcal{L}_{\mathrm{eff}}=\frac{1}{2}\mu\overline{\psi}\sigma_{\mu\nu}%
\psi(F^{\mu\nu}-\tan\theta_{W}Z^{\mu\nu}),
\end{equation}
where $F^{\mu\nu}$ is the field strength for photon, $Z^{\mu\nu}$ for $Z$
boson and $\mu$ is the magnetic dipole moment. The DM annihilations therefore
produce the standard model particles via $\gamma,Z$ exchanges. The
annihilation processes were studied in Ref. \cite{heo09} in detail, and here
we take advantage of the results (annihilation rates or fractions). The
corresponding annihilation fractions are tabulated in Table I for our
benchmark mass, 100 GeV.

\begin{table}[t]
\caption{Annihilation fractions for each channel, in which $u$ denotes up type
quarks $(=u,c)$, $d$ down type quarks $(=d,s,b)$, $\nu$ neutrinos $(=\nu
_{e},\nu_{\mu},\nu_{\tau})$ and $e$ charged leptons $(=e,\mu,\tau)$. Five
fundamental channels are considered for dark matter mass of 100 GeV.}%
\label{table1}%
\begin{ruledtabular}
\begin{tabular}{cccccc}
$\text{Channel}$ & $u\bar{u}$  & $d\bar{d}$& $\nu \bar{\nu}$ & $e\bar{e}$ &$W^+W^-$ \\
\hline
$\text{Annihilation fraction (\%)}$ & $14.6$ & $7.4$&$3.2$ & $10.1$ & $8.3$ \\
\end{tabular}
\end{ruledtabular}\end{table}

\section{Cosmic ray signatures}

DM may annihilate at some point in Galaxy and produce the standard model
particles. These produced particles then propagate in the interstellar medium.
Antimatter particles and photons have been considered to be the subject of
indirect DM searches, because antimatter particles are rarely produced in
astrophysical process and gamma rays can transport freely without energy loss
or transmutation of the direction. They may thus provide important signatures
of DM in the Galaxy. Observations of such signals can reveal information on
the microscopic nature of DM.

The emissivity/energy (production rate or source for the signals)\footnote{If
DM is produced with a primordial asymmetry like baryons, there would be almost
no signals from DM-antiDM annihilations due to lack of anti-dark matters
(antiDMs). This source is for equal populations of DM and antiDM, and hence
our predicted signals will be upper limits of the predictions. Recently a
mechanism of DM-antiDM oscillations is suggested to re-equilibrate the
populations at late times \cite{stu12, mci11}.} at location $\mathbf{x}$ from
the Galactic center is obtained from the convolution over the various
annihilation channels $f$ of the annihilation rate $\left\langle \sigma
v\right\rangle _{f}$ with the differential yield (single particle
spectra)\footnote{We use PYTHIA \cite{tor07}, as implemented in DarkSUSY
\cite{pgo04} or MicrOMEGAs \cite{gbe10} program, to generate the differential
yields (injected particle spectra).} $\left(  dN^{f}/dT\right)  _{a}$ for the
final state particles $a$,%
\begin{equation}
Q_{a}\left(  \mathbf{x},T\right)  =\frac{1}{4}B\left\langle \sigma
v\right\rangle _{f}\left(  \frac{dN^{f}}{dT}\right)  _{a}\left(  \frac
{\rho(\mathbf{x})}{M}\right)  ^{2},
\end{equation}
where $M$ is the DM mass, $B$ is an overall boost (enhancement) factor and
$\rho(\mathbf{x})$ is the DM mass density at the location $\mathbf{x}$. The DM
mass density around the Galactic center (DM halo profile) is not known,
especially near center ($\leq100$ pc$)$. The theoretically motivated ones are
Navarro-Frenk White (NFW) \cite{jfn97}, Moore \cite{bmo98}, cored Isothermal
\cite{jnb80}, and recently Einasto profiles \cite{jei65,jfn08}. The kinetic
energy $T$ is often approximated to the total energy $E$, in the case when the
particles are energetic. Notice that the factor $\frac{1}{4}$ is different
from the one of self annihilating DM.

The DM density profile can be parameterized as%
\begin{equation}
\rho(r)=\rho_{\odot}\left[  \frac{r_{\odot}}{r}\right]  ^{\gamma}\left[
\frac{1+(r_{\odot}/r_{s})^{\alpha}}{1+(r/r_{s})^{\alpha}}\right]  ^{\left(
\beta-\gamma\right)  /\alpha} ,
\end{equation}
where $\rho_{\odot}\simeq0.4$ GeV/cm$^{3}$ \cite{rca10} is the DM density in
the solar vicinity and $r_{\odot}=8.33$ kpc is the distance of the solar
system from the Galactic center. The profile parameters $\alpha,\beta
,\gamma,r_{s}$ are summarized in Table II. The Einasto profile is
$\rho_{\text{Einasto}}(r)$ =$\rho_{\odot}\exp\left[  -\left(  2/\alpha
)[(r/r_{s})^{\alpha}-1\right)  \right]  ]$ with $r_{s}=20$ kpc and
$\alpha=0.17$. As is well known, the NFW and Moore profiles exhibit a cusp at
the center of Galaxy.

\begin{table}[t]
\caption{The dark matter density parameters.}%
\begin{ruledtabular}
\begin{tabular}{cccccc}
$\text{Halo model}$ & $\alpha$  & $\beta$& $\gamma $ & $r_s$ \text {(kpc)}  \\
\hline
$\text{Navarro, Frenk, White}$ & $1$ & $3$&$1$ & $20$ \\
$\text{Moore}$ & $1.5$ & $3$&$1.5$ & $28$\\
$\text{cored Isothermal}$ & $2$ & $2$&$0$ & $5$  \\
\end{tabular}
\end{ruledtabular}\end{table}

The charged particles produced by the DM annihilation are predicted to come
from the halo near the Sun, not too far from the Sun at least, because they
may lose the energy while propagating through the Galactic halo. They are
deflected by the Galactic magnetic field, and this property has been described
by space diffusion \cite{fca02}. The charged particles suffer energy losses
from synchrotron radiation and inverse Compton scattering. The solar
modulation can also induce a certain amount of energy loss. Their energy
spectrum at the Earth, therefore, differs from the one produced at the source.
The equation that describes the evolution of the energy distribution for the
charged particles may be expressed as%
\begin{align}
\frac{\partial}{\partial t}\left(  \frac{dn}{dT}\right)  _{a}-\nabla
\cdot\left(  K\left(  T\right)  \nabla\left(  \frac{dn}{dT}\right)
_{a}\right)  -  &  \frac{\partial}{\partial T}\left(  b\left(  T\right)
\left(  \frac{dn}{dT}\right)  _{a}\right)  +\frac{\partial}{\partial z}\left(
sign(z)V_{C}\left(  \frac{dn}{dT}\right)  _{a}\right) \nonumber\\
&  =Q_{a}\left(  \mathbf{x},T\right)  -2h\Gamma_{ann}\delta\left(  z\right)
\left(  \frac{dn}{dT}\right)  _{a},
\end{align}
where $dn/dT$ is the number density of particles per unit volume and energy.
The second term accounts for the space diffusion with the energy dependent
diffusion constant $K\left(  T\right)  =K_{0}\left(  T/\text{GeV}\right)  $.
The energy loss due to synchrotron radiation in the Galactic magnetic field
and inverse Compton scattering on CMB photons and on Galactic starlight is
described in the third term. The rate of energy loss is $b\left(  T\right)
=T^{2}/($GeV $\tau_{T})$, where $\tau_{T}=10^{16}$ s is the energy loss time.
The fourth term is the effect of convective wind. The last term accounts for
the annihilation of the produced matter(s) in the interstellar medium, H and
He atoms, with annihilation rate $\Gamma_{ann}$ in the disk of thickness
$2h\simeq0.2$ kpc, and hence it is provided as a negative source term. The
relevant coefficients were parameterized, and the established (transport)
parameters were estimated from the analysis of observed isotope ratios in
cosmic rays, primarily the boron to carbon (B/C) ratio \cite{fdo04}. Three
propagation models have been featured with the established parameters, and
these propagation models correspond to minimal (MIN), medium (MED) and maximum
(MAX) antiproton fluxes \cite{fdo04,tde08}.

The number density $dn/dT$ is obtained by solving Eq. (4) with the steady
state condition $\frac{\partial}{\partial t}\left(  \frac{dn}{dT}\right)
_{a}=0$ and boundary conditions in a two-zone model \cite{fdo01}, where the
region of diffusion of cosmic rays is represented by a thick disk of thickness
$2L\simeq5-20$ kpc and radius $R\simeq20$ kpc, and the thin Galactic disk lies
in the middle and has thickness $2h$, radius $R$. The boundary conditions are
such that the number density vanishes at $z=\pm L$ and at $r=R$.

\subsection{The positron Channels}

The energy spectrum of positrons is obtained by solving the diffusion
equation, keeping only contributions of space diffusion and energy losses,%
\begin{equation}
-K\left(  E\right)  \nabla^{2}\left(  \frac{dn}{dE}\right)  _{e^{+}}%
-\frac{\partial}{\partial E}\left(  b\left(  E\right)  \left(  \frac{dn}%
{dE}\right)  _{e^{+}}\right)  =Q_{e^{+}}\left(  \mathbf{x},E\right)  ,
\end{equation}
with the relevant parameters listed in Table III. This diffusion equation may
be solved by the Green function formalism or the Bessel-transform method, and
the solution results in the following form:%
\begin{equation}
\left(  \frac{dn}{dE}\right)  _{e^{+}}=\frac{1}{b\left(  E\right)  }\int
_{E}^{M}dE_{S}Q_{e^{+}}\left(  \mathbf{x}_{\odot},E_{S}\right)  I_{e^{+}%
}(E,E_{S}),
\end{equation}
where $\mathbf{x}_{\odot}$ is the location of the Sun from the Galactic
center. The function $I_{e^{+}}(E,E_{S})$ must fully encode the Galactic
astrophysics from the input energy $E_{S}$ to energy $E(\leq E_{S})$, and the
full expression of this function can be found in Ref. \cite{tde08}.

\begin{table}[t]
\caption{Typical diffusion parameters for positrons deduced from a variety of
cosmic ray data, which yield the minimum (MIN), median (MED) and maximal (MAX)
fluxes.}%
\begin{ruledtabular}
\begin{tabular}{ccccc}
$\text{Model}$ & $\delta$  & $K_0 \text{[kpc$^2$/Myr]}$& $L \text{(kpc)}$ \\
\hline
$\text{MIN}$ & $0.55$ & $0.00595$& $1$  \\
$\text{MED}$ & $0.70$ & $0.0112$& $4$ \\
$\text{MAX}$ & $0.46$ & $0.0765$& $15$ \\
\end{tabular}
\end{ruledtabular}\end{table}

The positron flux is then given by%
\begin{equation}
\phi_{e^{+}}=B\frac{v_{e^{+}}}{4\pi}\left(  \frac{dn}{dE}\right)  _{e^{+}%
}=B\frac{v_{e^{+}}\left\langle \sigma v\right\rangle _{f}}{16\pi b\left(
E\right)  }\left(  \frac{\rho_{\odot}}{M}\right)  ^{2}\int_{E}^{M}%
dE_{S}\left(  \frac{dN^{f}}{dE}\right)  _{e^{+}}I_{e^{+}}(E,E_{S}),
\end{equation}
where $v_{e^{+}}$ is the velocity of the positron.

The enhancement may come from subhalo structure (dark clumps), and the
non-perturbative Sommerfeld effect because one of the force carriers is photon
in this scenario. The enhancement by the Sommerfeld effect can be calculated
from the original form \cite{aso31}. We can split the dipole operator in
energy and momentum dependent parts by the familiar Gordon decomposition,%
\begin{equation}
\mu\overline{v}(p^{\prime})\sigma^{\mu\nu}q_{\nu}u(p)=i\mu\overline
{v}(p^{\prime})(2M\gamma^{\mu}+(p^{\prime}-p)^{\mu})u(p).
\end{equation}
In the energy dependent part, we have the same type of coupling with the DM
mass dependence as for electric charge. The original Sommerfeld enhancement
factor is%
\begin{equation}
S=\frac{\pi\alpha_{\mu}/v}{1-e^{-\pi\alpha_{\mu}/v}}\overset{\alpha_{\mu}\gg
v}{\sim}\frac{\pi\alpha_{\mu}}{v},
\end{equation}
where $v$ is the DM velocity $(\sim10^{-3})$. In the original form,
$\alpha_{\mu}$ is the electric fine structure constant, but $\alpha_{\mu
}=16\pi\mu^{2}M^{2}$ in this case. We do not find series of resonances because
the potential is not localized. We have the enhancement $S\simeq16$ for the DM
mass of 100 GeV and magnetic dipole $\mu\simeq0.1$ TeV$^{-1}$. According to
the recent work \cite{Trs09,Sga09}, DMs annihilating after recombination may
contribute to the CMB anisotropy spectrum, and the enhancement bound could be
set up. The CMB bound is $S<\left(  120/f\right)  \left(  M/\text{TeV}\right)
$ where the parameter $f$ indicates the average fraction of energy absorbed by
the gas and depends on final states. The bound is slightly over $17$ for
$e^{\pm}$ final state, $f\simeq0.7.$ Our enhancement factor lies very near the
estimated bound. We can also consider an enhancement from the metastable bound
state between DMs, called ``WIMPonium". This production process from DMs
annihilation is, however, kinematically forbidden, because the estimated
kinetic energy $(\sim Mv^{2})$ with velocity $v$ $\sim10^{-3}$ is too small to
incorporate the binding energy $(\sim M\alpha_{\mu}^{2})$.

The strength of magnetic dipole is, according to Ref. \cite{heo09}, almost
constant for the DM mass larger than 100 GeV, and thereby the enhancement
increases in the square of the DM mass. Due to this reason, the fluxes, Eq.
(7), must have the similar magnitude about the DM masses larger than 100 GeV
because the fluxes are not scaled by the DM mass. We notice that the fluxes
are scaled by the DM mass in most of the models, and large boost factors have
been required for larger DM mass because the fluxes decrease with the square
of the DM mass.%

\begin{figure}
[ptb]
\begin{center}
\includegraphics[
height=4.1641in,
width=6.1834in
]%
{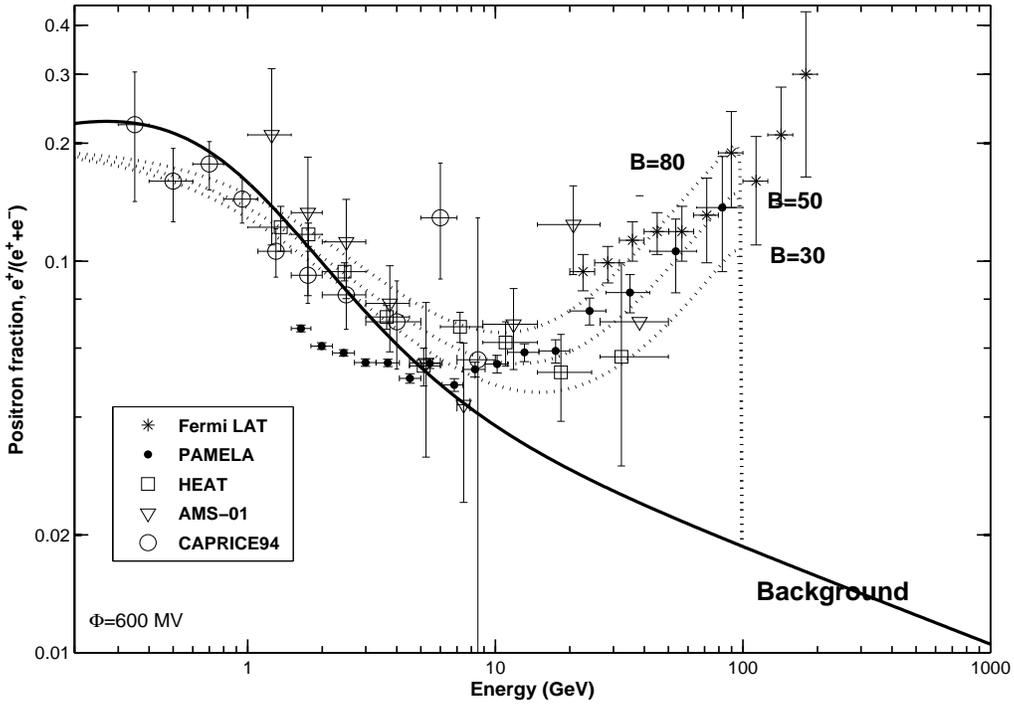}%
\caption{Positron fraction from the annihilation of the fermionic dark matter
particle. The boost factors have been chosen to provide qualitatively good
fits to the data with dark matter mass 100 GeV. Shown are the background in
solid-line and the experimental datasets of the PAMELA \cite{oad09}, HEAT
\cite{swb97}, AMS-01 \cite{mag07}, Fermi LAT \cite{mac12}, and CAPRICES94
\cite{mbo00}.}%
\end{center}
\end{figure}

The positrons are affected by solar wind and lose energy while transporting in
the solar system. This effect leads to a shift in the energy distribution
between the interstellar spectrum (IS) and the spectrum at the top of the
atmosphere (TOA). This modulation is considered for the predicted fluxes by
the relation in force field approximation \cite{ljg68},%
\begin{equation}
\phi_{e^{+}}^{\text{TOA}}(E)=\frac{E^{2}-m_{e^{+}}^{2}}{\left(  E+\left\vert
Z\right\vert e\Phi\right)  ^{2}-m_{e^{+}}^{2}}\phi_{e^{+}}^{\text{IS}%
}(E+\left\vert Z\right\vert e\Phi),
\end{equation}
where $\left\vert Z\right\vert $ is the magnitude of electric charge (1 in
this case), $e$ is the electric constant and $\Phi$ is the Fisk potential,
namely solar modulation parameter, which varies between 500 MV and 1.2 GV over
the eleven-year solar cycle. Since experiments are usually undertaken near
solar minimum activity, we choose $\Phi=600$ MV (the Fisk potential for the
PAMELA experiment) for our numerical analysis.

We show in Fig. 1 the predicted positron fractions for the boost factors,
$B=30,50,80$, with the computed background\footnote{Secondary positrons from
nuclear interactions of cosmic ray nuclei with interstellar gas have been
investigated in detail by the authors of Ref. \cite{ivm98}. Recently, it has
been suggested that the secondary positrons are increased by up to 60 per cent
at high energies above 100 GeV \cite{jla11}, based on analysis of the spectral
hardening in the cosmic ray proton and helium fluxes reported by the ATIC2
\cite{pan09} and CREAM \cite{ahn10} balloon experiments.} from Ref.
\cite{ivm98} and several experimental data sets. The predicted fractions have
almost no difference\footnote{If we only consider the fractions or fluxes of
signals excluding the background, there are sizable differences at the low
energies less than 10 GeV, especially in diffusion models} in the halo
profiles or the diffusion models. The predicted fractions exhibit a rather
sharp distribution at $E_{e^{+}}\simeq M$, since our candidate can directly
annihilate into electron and positron pair. The PAMELA \cite{oad09} have shown
a steep rise in the $10-100$ GeV range in their measurements and confirmed the
results of HEAT \cite{swb97} and AMS-01 \cite{mag07} experiments. Recently,
the steep rise has been extended to 200 GeV with three more data points over
100 GeV at the Fermi LAT \cite{mac12}. The predicted signals with the boost
factor $30-80$ nicely fit measurements of the PAMELA for the DM mass of around
100 GeV. An enhancement of about a factor of 16 comes from the Sommerfeld
effect, and the rest, an enhancement factor of $2-5,$ is expected to come from
subhalo structure (dark clumps). The existence of subhalos is a generic
prediction of the $\Lambda$CDM scenario of structure formation in the
Universe, and high resolution simulations \cite{jdi08} show that the large
scale structures form by continuous merging of smaller hallos which could be
in the form of subhalos. The contribution of subhalos to the flux could be
constrained from analysis of CMB data which do not rely on uncertain
assumptions of the DM distribution \cite{lpi11}. The subhalo boost factor has
been predicted to be ten \cite{lpi11} at most. Extracting the accurate
formalism of this boost factor is out of scope of this paper. Our subhalo
boost factors must be in the reasonable range.

We also have the difference between predictions and experimental measurements
or background at low energies ($\leq10$ GeV). It has been noticed that solar
modulation effect we consider has no charge-sign dependence and has to be
modified. This must be a future study.

\subsection{The antiproton channels}

The propagation of antiprotons is dominated by diffusion and the effect of the
Galactic wind. The energy spectrum of antiprotons is obtained by solving the
following diffusion equation:
\begin{equation}
\frac{\partial}{\partial z}\left(  sign(z)V_{C}\left(  \frac{dn}{dT}\right)
_{\overline{p}}\right)  -\nabla\cdot\left(  K\left(  T\right)  \nabla\left(
\frac{dn}{dT}\right)  _{\overline{p}}\right)  =Q_{a}\left(  \mathbf{x}%
,T\right)  -2h\Gamma_{ann}\delta\left(  z\right)  \left(  \frac{dn}%
{dT}\right)  _{\overline{p}},
\end{equation}
with the relevant parameters listed in Table IV. An important difference with
the positron case is that energy loss of antiprotons is negligible, because
antiprotons are more massive and hence it is absent in the diffusion equation (11).

\begin{table}[t]
\caption{Typical diffusion parameters for anti-protons deduced from a variety
of cosmic ray data, that yield the minimum (MIN), median (MED) and maximal
(MAX) fluxes.}%
\begin{ruledtabular}
\begin{tabular}{ccccc}
$\text{Model}$ & $\delta$  & $K_0 \text{[kpc$^2$/Myr]}$& $V_C \text{(km/s)}$ & $L \text{(kpc)}$ \\
\hline
$\text{MIN}$ & $0.85$ & $0.0016$&$13.5$ & $1$  \\
$\text{MED}$ & $0.70$ & $0.0112$&$12$ & $4$ \\
$\text{MAX}$ & $0.46$ & $0.0765$&$5$ & $15$ \\
\end{tabular}
\end{ruledtabular}\end{table}

The antiproton flux is then given by%
\begin{equation}
\phi_{_{\overline{p}}}=B\frac{v_{_{\overline{p}}}}{4\pi}\left(  \frac{dn}%
{dT}\right)  _{\overline{p}}=B\frac{v_{\overline{p}}\left\langle \sigma
v\right\rangle _{f}}{16\pi}\left(  \frac{dN^{f}}{dT}\right)  _{\overline{p}%
}\left(  \frac{\rho_{\odot}}{M}\right)  ^{2}I_{\overline{p}}\left(  T\right)
,
\end{equation}
where the function $I_{\overline{p}}\left(  T\right)  $ encodes all the
astrophysics, and the full expression can be found in Refs. \cite{pch96,lbe99}.%

\begin{figure}
[ptb]
\begin{center}
\includegraphics[
height=4.1641in,
width=6.1834in
]%
{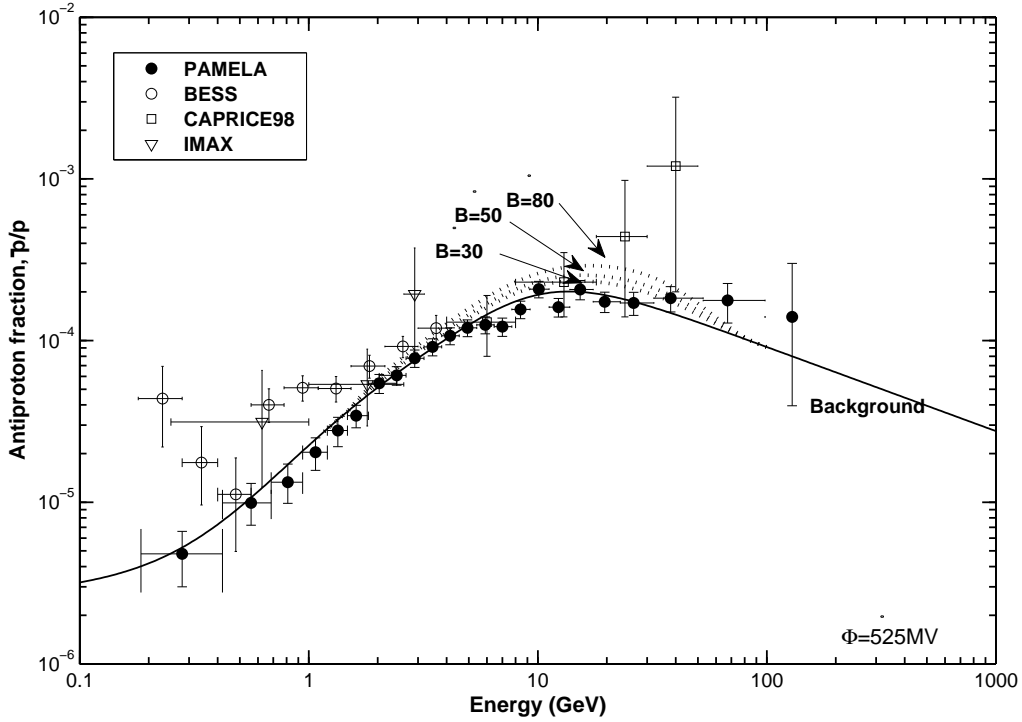}%
\caption{Antiproton over proton ratio as a function of kinetic energy. Shown
are the background in solid-line and the experimental data sets of the PAMELA
\cite{oad10}, BESS \cite{yas02}, CAPRICE98 \cite{mbo01} and IMAX \cite{jwm96}.
}%
\end{center}
\end{figure}

Fig. 2 shows the predicted ratios of antiproton over proton with the computed
background \cite{aml05} and several experimental measurements. As in the case
of the positrons, the predictions have almost no difference in the halo
profiles, but they are sensitive to the propagation models. The predictions in
the MIN propagation model are selected, and they may be within the invisible
range for current detectors. However, this scenario is most likely ruled out
for other propagation models\footnote{To be strict, the MED propagation model
can also be viable because of the uncertainty of the transport parameters. The
parameters are established by the best fit of cosmic ray B/C data
\cite{fdo04}. However, the assigned values of transport parameters may differ
by one order of magnitude.}, MED and MAX, because the predicted fluxes in MED
or MAX propagation model are about ten or one hundred times larger than the
ones in the MIN propagation model.

\subsection{The gamma-ray channels}

The production of gamma-rays has been considered to be a very important
channel to search for the DM signals, since they travel in straight lines and
can travel greater distances without energy loss. For these reasons, they
contain spectral and directional information that can be well measured. The
gamma-ray flux from the DM annihilations at a given photon energy from a
direction that forms an angle $\psi$ between the direction of the Galactic
center and that of observation is accounted for by the line-of-sight (los)
integration method,%
\begin{equation}
\phi_{\gamma}=B\frac{\left\langle \sigma v\right\rangle _{f}}{16\pi M^{2}%
}\left(  \frac{dN^{f}}{dE}\right)  _{\gamma}\int_{\text{los}}\rho^{2}%
(r(s,\psi))ds,
\end{equation}
where $r=\sqrt{r_{\odot}^{2}+s^{2}-2r_{\odot}s\cos\psi}$ is the Galactocentric
distance. In terms of the galactic latitude $b$ and longitude $l$, one has
$\cos\psi=\cos b$ $\cos l$. The coordinate $s$ parameterizes the distance from
the Sun along the los.

This form can be reduced to%
\begin{equation}
\phi_{\gamma}\simeq8.31\times10^{-11}\text{ }\left(  \text{cm}^{2}%
\cdot\text{sr}\cdot\text{s}\right)  ^{-1}\cdot\frac{B\left\langle \sigma
v\right\rangle _{f}}{10^{-26}\text{ cm}^{3}\text{s}^{-1}}\left(
\frac{100\text{ GeV}}{M}\right)  ^{2}\left(  \frac{dN^{f}}{dE}\right)
_{\gamma}\cdot\overline{J}_{\Delta\Omega}~,
\end{equation}
where $\overline{J}_{\Delta\Omega}$ is a dimensionless los integral averaged
over the solid angle $\Delta\Omega$ and is defined by%
\begin{equation}
\overline{J}_{\Delta\Omega}=\frac{1}{\Delta\Omega}\int J\left(  \psi\right)
d\Omega~,
\end{equation}
with%
\begin{equation}
J\left(  \psi\right)  =\int\frac{\rho^{2}(r)}{\rho_{\odot}^{2}}\frac
{ds}{r_{\odot}}~.
\end{equation}
%

\begin{figure}
[ptb]
\begin{center}
\includegraphics[
height=1.2021in,
width=4.9692in
]%
{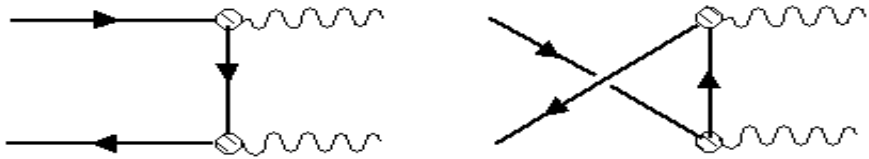}%
\caption{Feynman diagrams for annihilation of dark matter to one-photon or
two-photons. The hatched circles indicate the dipole couplings.}%
\end{center}
\end{figure}

In addition to the continuum emission, the direct DM annihilations produce
$\gamma\gamma$ and $\gamma Z$ final states\footnote{The production of the
single $\gamma$ or $Z$ final state is prohibited, because in this process it
is impossible to conserve energy and momentum together.} in this scenario, in
which Feynman diagrams are shown in Fig. 3. Such processes would yield the
very distinctive feature of monoenergetic gamma-ray lines (monochromatic
photons) with an energy $E_{\gamma}=M$ or $M\left(  1-m_{Z}^{2}/(2M)^{2}%
\right)  $. The full annihilation rates for the production of $\gamma\gamma$
are $\left\langle \sigma v\right\rangle _{\gamma\gamma}=\mu^{4}M^{2}/2\pi$ and
$\left\langle \sigma v\right\rangle _{\gamma Z}=\mu^{4}M^{2}\beta_{Z}^{2}%
\tan^{2}\theta_{W}/4\pi$ for $\gamma Z.$ The contributions for $\gamma\gamma$
and $\gamma Z$ final states are suppressed by the magnetic dipole $\mu^{4}$,
and the $\gamma Z$ final state has an additional suppression with Weinberg
angle $\tan^{2}\theta_{W}$.

We restrict our analysis to the possible signals from the Galactic halo for a
complementarity in the case of the positrons, and the same boost
factors\footnote{It has been predicted that the possible enhancement from
subhalo has an angular and/or energy dependence on the cosmic rays. The
enhancement can be different in each channel, but there are still no clear
experimental evidences for dark clumps. We select the same boost factors,
$B=30-80,$ for a reference. The main idea of the enhancement from subhalo
structure comes from $\left\langle \rho^{2}\right\rangle \geq\left\langle
\rho\right\rangle ^{2}$.} $B=30,50,80$ are selected. We compare our
predictions to experimental observations in two stringent cases. One is a
process that may contribute to the extragalactic gamma-ray background (EGB).
The other is, in the case that astrophysical sources account for the EGB in
the entire energy range, a process which satisfies the experimental exclusion
limit of the Fermi LAT \cite{aaa10}.

Fig. 4 shows the predicted gamma-ray spectra as a function of photon energy in
the region $0^{\circ}\leq\ell\leq360^{\circ},\left\vert b\right\vert
\geq10^{\circ}$. The spectra are superpositions of the continuum and
monoenergetic gamma-rays at the DM mass of 100 GeV. The spectra have almost no
difference in the halo profiles. The EGB from the Fermi LAT \cite{aaa10} is
given by
\begin{equation}
E_{\gamma}^{2}\phi_{\gamma}\simeq5.5\times10^{-7}\left(  \frac{E}{1\text{GeV}%
}\right)  ^{-0.41}\text{ }\left(  \text{cm}^{2}\cdot\text{sr}\cdot
\text{s}\right)  ^{-1}\text{GeV.}%
\end{equation}
The same type of background from the analysis of the EGRET measurements is
also described \cite{awst04} from the first analysis \cite{psr98}. The
predicted spectra are not exceeding the EGB, and we might have a signature if
it can be disentangled from astrophysical ones. In addition, we check if our
prediction can account for the EGRET anomaly\footnote{Although this anomaly is
likely caused by a systematic error of the effective detector sensitivity
calibration \cite{fws08}, we include this anomaly in our analysis for a
possible signal in case.} which is not confirmed at the Fermi LAT. Our
predictions are too soft to explain the observation of the EGB, and the more
enhancement would be needed.%

\begin{figure}
[ptb]
\begin{center}
\includegraphics[
trim=0.000000in 0.000000in 0.000000in -0.138416in,
height=3.5129in,
width=4.8179in
]%
{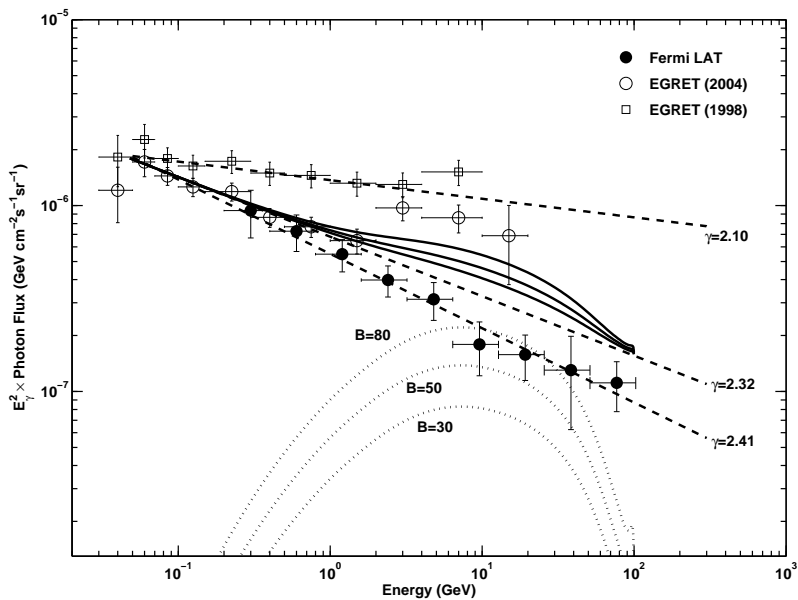}%
\caption{The gamma-ray spectrum from Galatic halo for dark matter mass of 100
GeV and the region, $0^{\circ}\leq\ell\leq360^{\circ},\left\vert b\right\vert
\geq10^{\circ}$. Shown are the experimental data sets from the Fermi LAT and
EGRET with their fitted spectral indices.}%
\end{center}
\end{figure}

The predicted fluxes have to be within the uncertainty of Fermi LAT data, in
the case that the EGB is accounted for by astrophysical sources in the entire
energy range. Fig. 5 shows the predicted fluxes with 90\%, 95\% and 99\% C.L.
experimental limits of the Fermi LAT, which are estimated from the data table
in Ref. \cite{aaa10}. The predictions are smaller than the exclusion limits,
and so satisfy the current experimental constraint.%

\begin{figure}
[ptb]
\begin{center}
\includegraphics[
trim=0.000000in 0.000000in 0.000000in -0.114324in,
height=3.5129in,
width=4.8179in
]%
{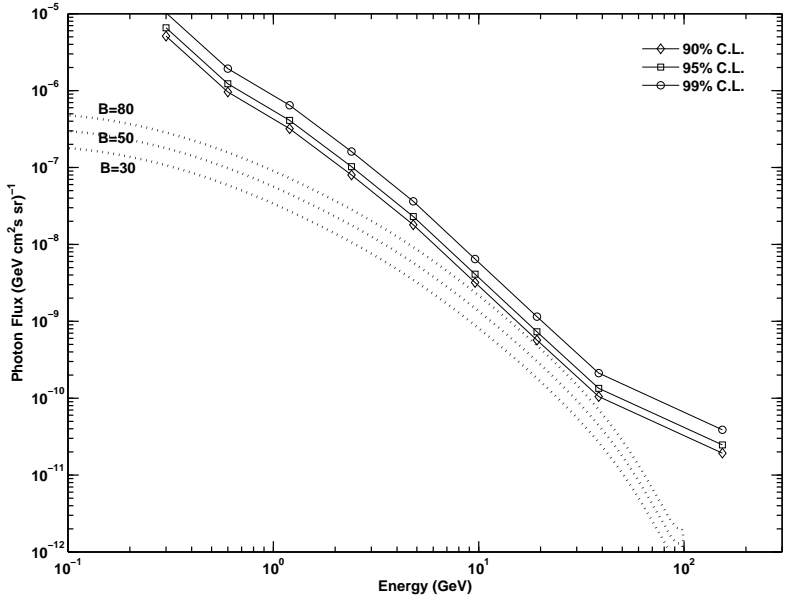}%
\caption{The gamma-ray flux from Galatic halo for dark matter mass of 100 GeV
and the target region $0^{\circ}\leq\ell\leq360^{\circ},\left\vert
b\right\vert \geq10^{\circ}$ with 90\%, 95\% and 99\% C.L. experimental limits
of the Fermi LAT.}%
\end{center}
\end{figure}

The Galactic center or the region close to it must be the most complex region
in the Galaxy due to many possible sources and the difficulty to model the
diffuse emission. Hence, it may be very difficult to disentangle possible DM
annihilation signals from the background fluxes. The monochromatic gamma-ray
lines appearing from DM annihilations could provide smoking-gun signatures for
these regions, because the line signals mostly cannot be mistaken for
astrophysical source.

Fig. 6 shows the predicted annihilation rate for $\gamma\gamma$ and $\gamma Z$
final states as a function of the DM mass. The Sommerfeld effect is included,
but not the subhalo structure. The strength of DM magnetic dipole is chosen to
satisfy the relic density from Ref. \cite{heo09}. The curves given for the
NFW, Einasto, and Isothermal DM distributions are 95\% C.L. upper limits of
Fermi LAT \cite{mack12} for the region of $\left\vert b\right\vert
\geq10^{\circ}$ plus a $20^{\circ}\times20^{\circ}$ square centered at the
Galactic center. The predicted annihilation rates are approximately
$4.0\times$ $10^{-29}$ cm$^{3}/$s at $E_{\gamma}\simeq100$ GeV for the
$\gamma\gamma$ final state and $4.0\times10^{-30}$ cm$^{3}/$s for the $\gamma
Z$ final state at $E_{\gamma}\simeq80$ GeV. Otherwise, the estimated upper
limits at the Fermi LAT are $10^{-27}-10^{-26}$ cm$^{3}/$s for both final
states, depending on the DM mass and the halo profiles. The predictions are
two or three orders smaller than the experimental upper limits. However, our
predictions can be enhanced if dense DM clumps are considered in regions close
to the Galactic center. Recently, the authors of Refs. \cite{cwe12,ete12}
pointed out the gamma-ray excess, $1-3\times10^{-27}$cm$^{3}$/s, around 130
GeV in the spectrum based on these measurements with 4.5 or 6$\sigma$
statistical significance. Our predictions with the subhalo boost factor of
about 100 can account for the gamma-ray excess.%

\begin{figure}
[ptb]
\begin{center}
\includegraphics[
trim=0.000000in 0.000000in -0.178258in 0.000000in,
height=4.1658in,
width=6.1834in
]%
{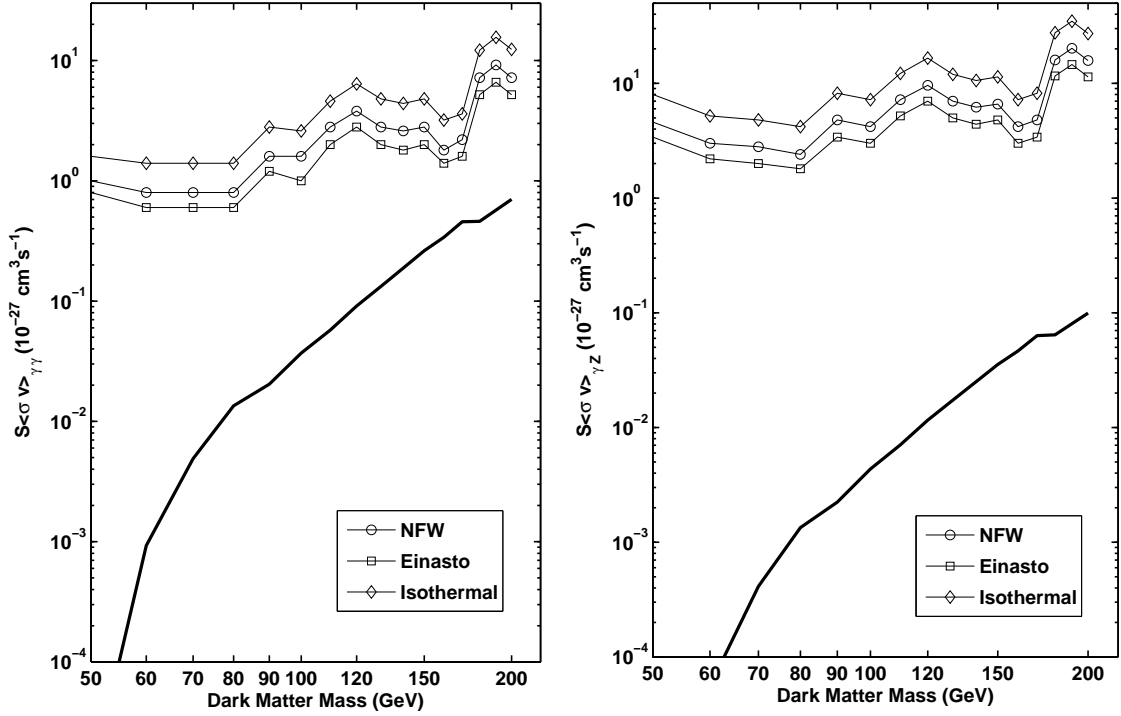}%
\caption{Annihilation rates with Sommerfeld enhancement factor $S$ for dark
matter annihilation to $\gamma\gamma$ or $\gamma Z$. Shown are the 95\% C.L.
upper limits of the Fermi LAT for the region of $\left\vert b\right\vert
\geq10^{\circ}$ plus a $20^{\circ}\times20^{\circ}$ square centered at the
Galactic center.}%
\end{center}
\end{figure}

One of the reasons we mostly get the low predictions for sharp peaks is due to
the relatively poor energy resolution. The current energy resolution of the
Fermi LAT is $11-13\%$ \cite{mack12} in the full width at half maximum. The
energy resolution could be as good as $1.5-2\%$ for a planned experiment,
AMS-02 \cite{ako10}, in which the upper limits of annihilation rate will be
$10^{-30}-10^{-29}$ cm$^{3}/$s. Our predicted signals are in the potential
probe at the AMS-02.

\section{Conclusion}

We considered cosmic ray signals in the positron, antiproton and photon
channels of dipole-interacting DM annihilation. The predicted signals in the
positron channel could nicely account for the excess of positron fraction from
Fermi LAT, PAMELA, HEAT and AMS-01 experiments for the DM mass larger than 100
GeV with a boost factor of $30-80$. An enhancement of about a factor of 16
could come from the Sommerfeld non-perturbative effect and the rest, an
enhancement factor of $2-5$, from subhalo structure (dark clumps). The
predicted signals have almost no dependence on the DM mass because of the
Sommerfeld effect. No excess of antiproton over proton ratio at the
experiments also gives a severe restriction for our scenario. This scenario
may be viable for MIN Galactic propagation model, but likely ruled out for
other propagation models, MED and MAX. The predicted signals from the Galactic
halo in the region $0^{\circ}\leq\ell\leq360^{\circ},\left\vert b\right\vert
\geq10^{\circ},$ and signals as the monoenergetic gamma-ray lines
(monochromatic photons) for the region ($\left\vert b\right\vert \geq
10^{\circ}$ plus a $20^{\circ}\times20^{\circ}$ square centered at the
Galactic center) close to the Galactic center were also considered. The
predicted signals from the Galactic halo must satisfy the current experimental
constraint, and the signals for the region near the Galactic center as
monoenergetic lines must be smaller than the experimental exclusion limits of
the Fermi LAT. The gamma-ray excess $1-3\times10^{-27}$cm$^{3}$/s around 130
GeV, pointed out by the authors of Refs. \cite{cwe12,ete12}, could be
accounted for in this scenario, with the subhalo boost factor of about 100.
Our predicted signals as monoenergetic lines for the region near Galactic
center are also in the potential probe at the planned experiment, AMS-02, with
the better experimental method.

\begin{acknowledgments}
The work was supported by the National Research Foundation of Korea (NRF)
grant funded by Korea government of the Ministry of Education, Science and
Technology (MEST) (No. 2011-0017430) and (No. 2011-0020333).
\end{acknowledgments}

\end{document}